\documentclass[12pt]{article}
\usepackage{amsmath}
\usepackage{graphicx,psfrag,epsf}
\usepackage{enumerate}
\usepackage{natbib}
\usepackage[font=footnotesize, labelfont={sf,bf}, margin=1cm]{caption}
\usepackage{hyperref}

\newcommand{\blind}{0}

\begin{document}

\def\spacingset#1{\renewcommand{\baselinestretch}%
{#1}\small\normalsize} \spacingset{1}

\if0\blind
{
  \title{\bf Near-Balanced Incomplete Block Designs with An Application to Poster Competitions}
  \author{Xiaoyue Niu\\
    and \\
    James L. Rosenberger \\
    Department of Statistics, The Pennsylvania State University}
	\date{}
  \maketitle
} \fi

\if1\blind
{
  \bigskip
  \bigskip
  \bigskip
  \begin{center}
    {\LARGE\bf Near-Balanced Incomplete Block Designs, with An Application to Poster Competitions}
\end{center}
  \medskip
} \fi

\bigskip
\begin{abstract}
Judging scholarly posters creates a challenge to assign the judges efficiently. If there are many posters and few reviews per judge, the commonly used Balanced Incomplete Block Design is not a feasible option.  An additional challenge is an unknown number of judges before the event.  We propose two connected near-balanced incomplete block designs that both satisfy the requirements of our setting: one that generates a connected assignment and balances the treatments and another one that further balances pairs of treatments.  We describe both fixed and random effects models to estimate the population marginal means of the poster scores and rationalize the use of the random effects model.  We evaluate the estimation accuracy and efficiency, especially the winning chance of the truly best posters, of the two designs in comparison with a random assignment via simulation studies.  The two proposed designs both demonstrate accuracy and efficiency gains over the random assignment.
\end{abstract}

\noindent%
{\it Keywords: judging system, population marginal means, random effects model}  
\vfill

\newpage
\spacingset{1.45} 
\section{Introduction}
\label{sec:intro}
The Graduate School at our university requested a new system for judging and scoring posters in their annual poster competition. Every year over 200 students enter their posters into the contest. Faculty members, post-docs, and graduate students can sign up as judges for the contest. There are usually around 100 judges. Each judge is assigned five posters to review and score. Each poster must have at least one faculty judge. A logistical complication of organizing the event arises because some judges who originally signed up do not show up, and some judges just show up without signing up in advance. So the organizers only learn the exact number of judges on the day of the event.  Moreover, since a different set of judges review each of the posters, the organizers realize that simply taking the average of the scores each poster gets does not provide a fair evaluation. Therefore, they were seeking a way to make the judge assignments in advance and create a method to ``fairly'' evaluate and score the posters.

\section{Method}
\subsection{The challenge: no BIBD exists for this situation}
We propose framing the judge assignment as a design problem, where the posters are the treatments and the judges are the blocks. This creates an incomplete block design (IBD) since every judge will only review and score a subset of five of the posters. The optimal design of this type would be a Balanced Incomplete Block Design (BIBD) (see, e.g., \cite{montgomery}), in which every poster is judged an equal number of times, and every pair of posters needs to be judged by the same judge an equal number of times. The following two relationships summarize the above restrictions:
\begin{eqnarray}
tr&=&bk \label{eqn:block_size}\\
\lambda&=&\frac{r(k-1)}{t-1}, \label{eqn:balance}
\end{eqnarray}
where $b$ is the number of blocks (judges), $r$ is the number of replicates (reviews per poster), $k$ is the block size (number of reviews per judge), $t$ is the number of treatment levels (posters), and $\lambda$ is the number of blocks in which each pair of treatments appears together (number of times each pair of posters is reviewed by the same judge). Suppose we have 201 posters and each judge reviews five posters, $t=201, k=5$. From Equation \ref{eqn:balance}, even with the minimum $\lambda=1$, a BIBD would require each poster to be reviewed by 50 judges ($r=50$). From Equation \ref{eqn:block_size}, we would need $b=\frac{tr}{k}=\frac{201\times 50}{5}=2010$ judges, which is not feasible. Moreover, the number of judges, i.e. blocks, available is unknown in advance. 

There are alternatives with good properties when balance is not achievable, such as a partially balanced incomplete block design (\cite{bose52}) or the alpha-type of resolvable incomplete block designs (\cite{patterson76}). In our application, since all the posters should communicate research to a general audience, they cannot be partitioned into pre-determined categories (classes). Therefore a partially balanced incomplete block design, with pre-determined classes, is not appropriate. Due to the unknown number of judges (blocks), we can not partition the blocks into sets of replicates; thus the design is not resolvable (\cite{yates36}). Therefore the alpha-type of resolvable design is not suitable either.  An extensive search of the literature did not reveal an exact match of an existing design. Motivated by lack of guidelines on how judges should be selected or how the final scores should be calculated for competitions, in the following sections we propose two designs that can be applied to a general judging system, the corresponding algorithm for implementation, and statistical methods to calculate the final scores. 

\subsection{Goals of the Designs}
The goals of the judge assignment scheme can be summarized as follows:
\begin{enumerate}
\item Each judge reviews a fixed number ($k$) of distinct posters (five in this case).
\item There are two types of judges, faculty and non-faculty. Each poster needs to be reviewed by at least one faculty judge. 
\item Regardless of how many judges ($b$) eventually show up, the assignment of judges must be \textit{connected}. A connected design means that every poster can be connected to every other poster by a chain of pairs within blocks (\cite{bose49}). The design and any \textit{initial subset}, defined as $A_i=\{1,2,...,i\}$, for all $i\le b$, of assignments needs to be connected. A connected design provides more precise estimates of the posters' true quality scores. 
\item The number of times each poster is judged ($r$) should not differ by more than one so that we get nearly equal replicates.
\item (optional) Each Pair of posters $(h,i)$ is reviewed by the same judge at most once. 
\end{enumerate}

We can achieve connectivity (\#3) if for each new judge assignment, one of the posters assigned to the judge comes from those previously reviewed. Given the total number of posters ($t$) and number of posters each judge reviews ($k=5$ in this case), we can calculate the minimum number of judges required to guarantee connectivity, $b_{min}$, to be the smallest integer $\geq t/4$.  

We can give the first $b_{min}$ judge assignments to faculty judges to satisfy the requirement \#2 above. To satisfy \#4, the maximum number of times each poster can be reviewed by any judge is the smallest integer $\geq b_{min}*5/t$. Denote this maximum review number by $r_f$. 

\subsection{The Algorithm}
To generate any desired number of judge assignments with the above goals in mind, we propose two designs. The first design satisfies goals \#1 to \#5, and is referred to as the type 1 near-balanced design (NB1). The following algorithm implements the NB1 design: 
\begin{enumerate}
\item Randomly select 5 posters from the pool for the first judge. Keep track of the number of times each poster ($i$) gets reviewed ($r_i$). 
\item From the second judge assignment to the $b_{min}$th assignments, the first review of each judge is always selected from the posters that have been previously reviewed, with the selection probability proportional to $r_f - r_i$.
\item The other four reviews are randomly selected from the set of least reviewed posters. To be more specific, first randomly select from the posters that have not been reviewed (the ones that have $r_i = 0$). Stop if we get four. If not, move on to the ones that have been reviewed once (the ones that have $r_i = 1$). Randomly select the remaining needed number of posters from this subset. Continue in the same fashion if necessary to those with two reviews, etc. 
\item After each new random assignment is generated, check how many times each pair of posters is reviewed by the same judge. If any pair appears more than once, replace the last assignment with a new one until $\lambda_{ij} \le 1$ for all $(i,j)$ pairs.  
\item From the ($b_{min}+1$)th assignment, generate all five reviews according to Step 3 and 4 until the desired number of judge assignments have been made.

\end{enumerate}

To keep the algorithm efficient, if any assignment takes more than a certain number of attempts (e.g. 500) due to the check in step 4, which implies there might not be a solution given the existing assignment, the algorithm will stop and start over from the beginning. 

If we drop Step 4 in the above algorithm, we will satisfy goals \#1 through \#4. We call this design the type 2 near-balanced design (NB2). NB2 balances the number of reviews of each poster, while NB1 further balances number of reviews of each pair of posters. 

After knowing the number of posters, we can sequentially generate any number of judge assignments using the above scheme, preferably more than needed in case more judges show up without signing up. Since we generate the judge assignment sequentially, any initial subset of the poster assignments to judges will automatically satisfy the requirements. Therefore, no matter how many judges actually show up, as long as the assignments are handed to judges according to the order generated, the design is always efficient. However we need to have a minimum number of faculty judges ($b_{min}$). If this condition is not met, we would have to drop requirement \#2 above and give the faculty assignments to general judges. If more faculty show up, then they will be treated as general judges. 

The above algorithm runs very fast on a personal computer. With the average poster number of around 200, and about 100 judges, the computational time ranges from 0.2 (NB2) to 5 (NB1) seconds. In other settings, NB1 potentially can take much longer due to the check step.

\subsection{The Statistical Model}
For an IBD, in order to estimate and compare all treatments, the simple average of the review scores is not valid due to the judge differences. Instead, we should estimate the \textit{population marginal means} (\cite{searle80}) which take into consideration and adjust for the judge effects. The statistical model to estimate the population marginal means is as follows:
\begin{equation}
Y_{ij}=\mu + P_i + J_j + \epsilon_{ij}, 
\label{eqn:model}
\end{equation}
where $Y_{ij}$ is the score that poster $i$ gets from judge $j$, $\mu$ is an unknown parameter, $P_i$ is the $i$th poster effect and $J_j$ is the $j$th judge effect. The judge factor can be modeled as a fixed or random factor. See \cite{montgomery} for a detailed description of the intrablock analysis (fixed effects) and interblock analysis (random effects). The error term $\epsilon_{ij}$ comes from two sources: the random scoring error and the poster $\times$ judge interaction effects, i.e. how the criterion of a particular judge interacts with the quality characteristics of each poster. As discussed in \cite{nelder77} and \cite{robinson91}, the choice of model is determined by the nature of the questions and the properties of the blocking factors. In our example, we treat the judges as a random blocking factor for several reasons. First there are many judges (over 100), and the population marginal mean estimates are more accurate with a smaller standard error unless the block variance is much larger than the error variance (\cite{robinson91}). Secondly we are not interested in estimating the judge effects but simply accounting for them. Thirdly the judges are only a sample of faculty and post-docs on campus. Finally, even though we propose a design that guarantees connectivity, in reality, there is the chance that the assignments are disconnected due to some administrative mistakes. In this case, the fixed effects model will fail, but the random effects model will provide estimates of the poster marginal means. 

\section{Simulation Study}
To evaluate the performance of the proposed designs, we conduct simulation studies in a setting that mimics the real competition. We set the number of posters to be 200, the number of judges to be 100, and the number of awards to be 30. From past data, we find that the poster mean score is around 80, the poster standard deviation is about 7, the judge standard deviation is about 6, and the standard deviation of the random errors is about 7. We consider three designs, NB1, NB2, and a random design. In the random design, each of the first several judges randomly selects 5 posters from the pool of unreviewed posters until all of the posters are reviewed at least once. Then each of the later judges randomly selects 5 from all of the posters. This particular random design only guarantees each poster being reviewed at least once without the other properties of NB1 and NB2.

In each iteration of the simulation, we generate a $200 \times 100$ matrix of scores $Y_{ij}$ according to Equation \ref{eqn:model} with the parameters being $\mu = 80$, $P_i \sim N(0,7^2)$, $J_j \sim N(0,6^2)$, and $\epsilon_{ij} \sim N(0,7^2)$. The score matrix is fixed at each iteration. We define the ``true score'' of poster $i$ as $\mu + P_i$, since we are interested in how an infinite number of judges from a hypothetical population would score the posters rather than how the actual judges in the sample would judge the posters (the latter one will add a term of average judge effect $\bar J_{\cdot}$ to the poster score). The true poster scores are held constant for the three designs at each iteration. The three designs will select three different subsets of poster-judge combinations as their corresponding design matrices. We fit the random effects model to each of the three subsets and estimate the population marginal means of the posters. We run 1000 such iterations. Out of the 1000 iterations, the random design generates 27 assignments that are disconnected, thus a fixed effects model would fail to produce poster score estimates in those runs. This supports both the use of random effects model and the near-balanced designs that can guarantee connectivity.

\begin{figure}[!ht]
\hspace{-0.5cm}
\includegraphics[height=4in]{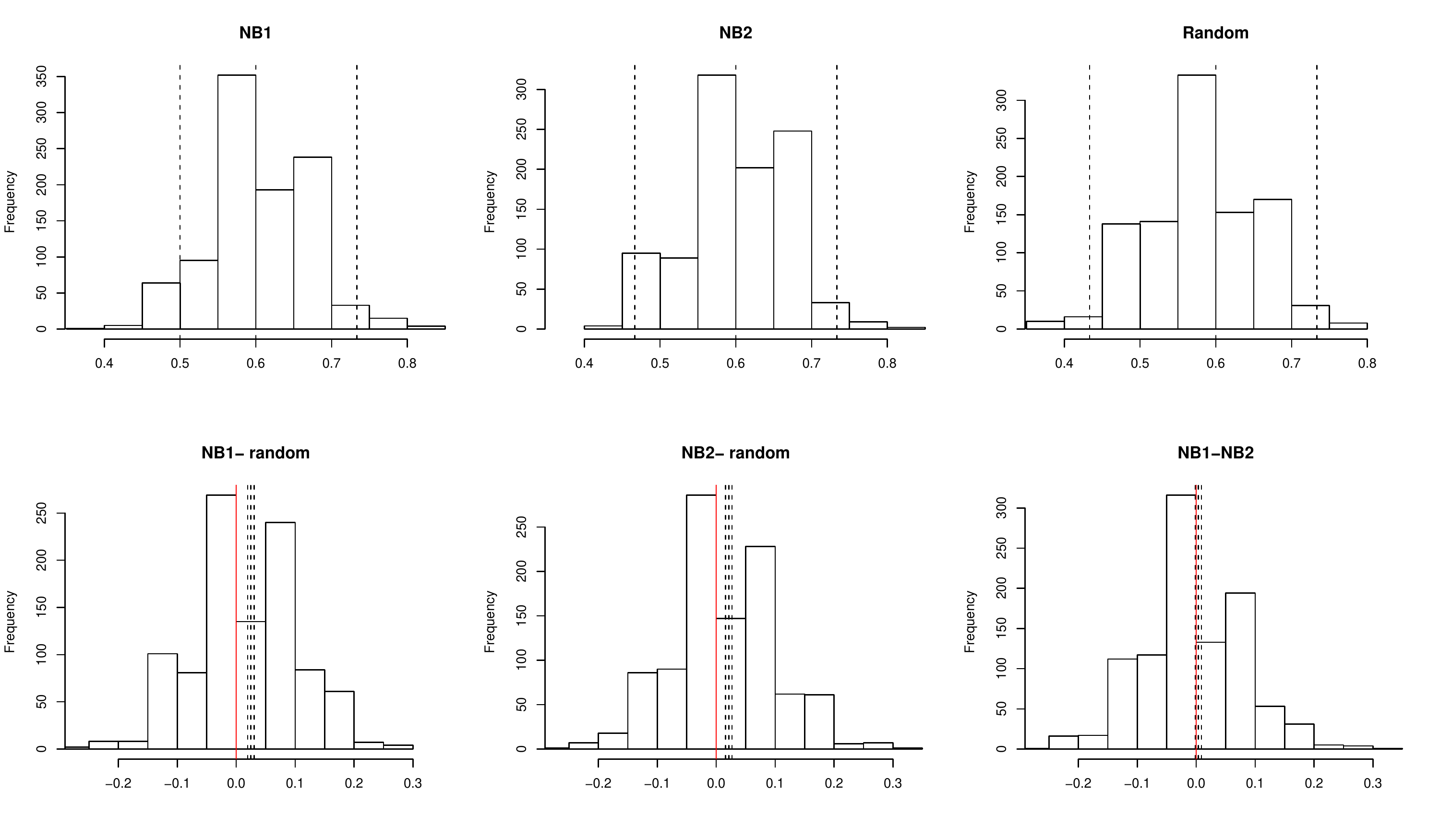}
\caption{Proportion of the 30 truly best posters judged to be in the top 30.  The top row summarizes the distribution of the proportions for the three designs (NB1, NB2, and random) across the 1000 simulation runs. The vertical lines indicate the 2.5\%, 50\%, and 97.5\% quantiles of each distribution. The bottom row summarizes the distribution of the differences of the proportions between each pair of designs. The solid vertical lines indicate 0. The two dashed vertical lines are the 95\% confidence bounds of the mean difference.}
\label{fig:win}
\end{figure}

Since the ultimate goal of the poster competition is to select the best posters to give awards, we set the probability of the ``truly best posters'' being selected as our primary evaluation statistic. The ``truly best posters'' are defined as the top 30 posters ranked by the true poster scores. Figure \ref{fig:win} examines, for each design, the proportion of the 30 truly best posters judged to be in the top 30.  Each histogram in the top row displays the empirical distribution of the proportion of the 30 truly best posters ranked among the top 30 across the 1000 simulation runs for a given design.  The median probability is 0.6 for all three designs, which implies that on average, 18 of the 30 truly best posters would win an award when awards are given to the 30 posters judged to be best. The 97.5\% quantile is 0.733 for all three designs, and the 2.5\% quantile is 0.5, 0.467, and 0.433 for NB1, NB2, and random respectively. Thus, we estimate that for 95\% of poster competitions matching our assumptions, 15-22 top posters are awarded under NB1, 14-22 under NB2, and 13-22 under random design. Note that the maximum probability is 0.833 for NB1 and NB2 and 0.8 for random, which implies that even under the best case scenario 5 of the best posters are not awarded. On average each poster gets only 2.5 reviews due to the limited number of reviews per judge. Therefore the estimated scores and ranks are highly variable. We think the fact that on average we only award 60\% of the truly best posters and never get all of the 30 completely correct is a common limitation in most of the judging systems. 

We further examine the difference of the three designs. In each iteration, we calculate the difference in the winning probabilities of all three pairs. The second row of Figure \ref{fig:win} displays the distribution of the differences. The dashed vertical lines indicate the mean difference and the 95\% confidence interval of the mean, with a solid vertical line at 0 as the reference. We can see that the mean difference between NB1 and random, and NB2 and random are both about 0.025 and are significantly greater than 0. The median difference is about 0.033 so the NB design gets 1 more poster correctly awarded on average. The relatively small difference is mainly due to the specific combination of poster number, judge number, and reviews per judge. In NB designs, half of the posters get 3 reviews and the other half get 2 reviews, while in the random design most of the posters still get 2 or more reviews and only a small number of posters get the extreme of a single review. In addition, the judges are sampled from a common distribution so they are relatively similar. Nonetheless, the difference is statistically significant so is not due to Monte Carlo error. Taken into consideration that the average winning probability is not high, we believe 1 poster difference is a scientifically meaningful difference. NB1 and NB2 seem to be equivalent in selecting the best posters. Due to the small difference in the number of reviews of each poster, the additional pair balance does not seem to increase the winning chance of the truly best posters. 

\begin{figure}[!ht]
\hspace{-0.5cm}
\includegraphics[height=4in]{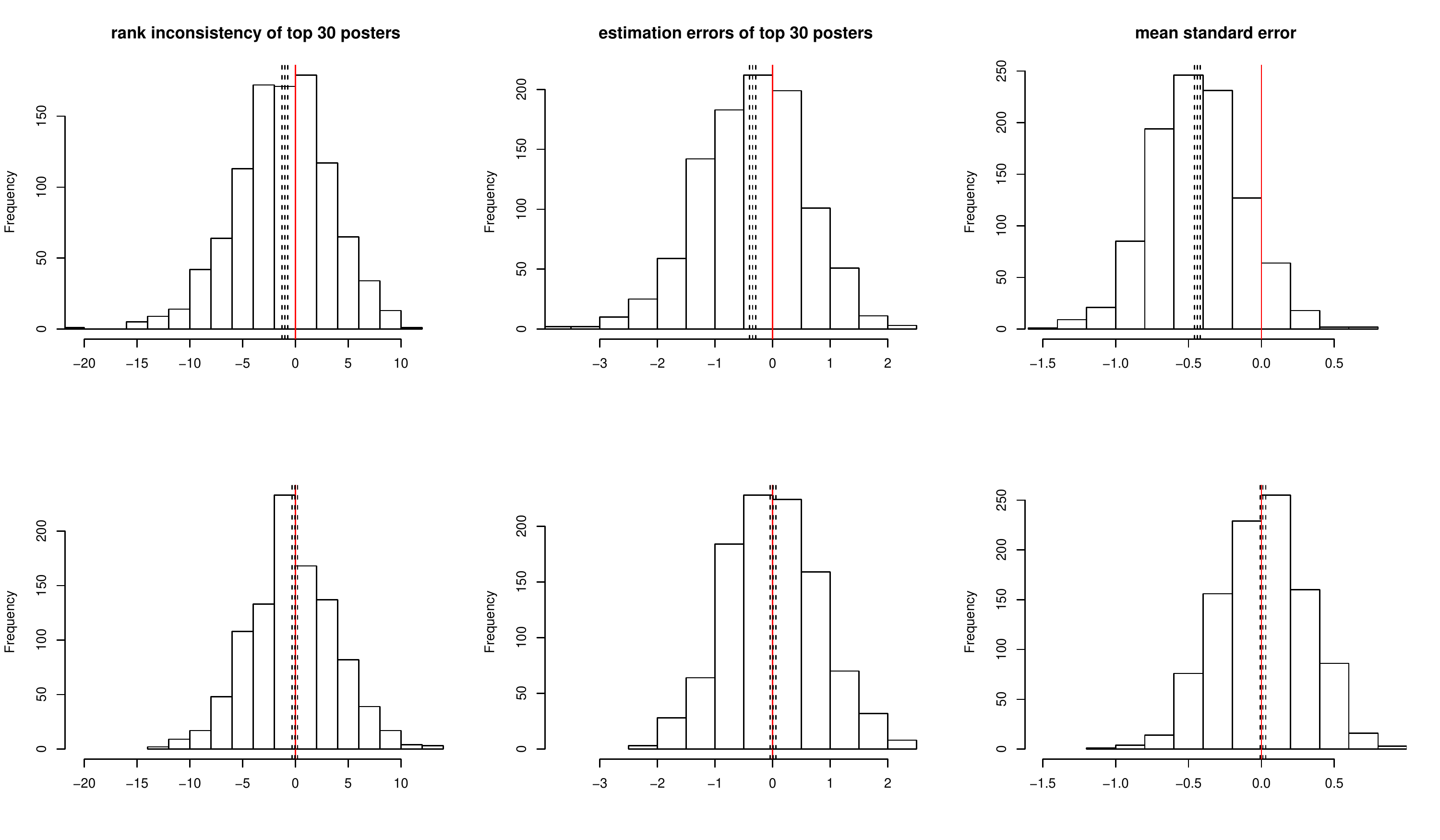}
\caption{Summary of the differences between NB1 and random assignment (top row) and NB1 and NB2 (bottom row) of three measures: median rank difference of the top 30 posters (first column), mean score deviance of the top 30 posters (2nd column), and mean standard error of the estimator (3rd column). The solid vertical line indicates 0. The dashed lines display the 95\% confidence bounds.}
\label{fig:other}
\end{figure}

In addition, we look at three other measures of the accuracy and efficiency of the designs: (1) the median of the absolute difference in the estimated and true ranks of the truly best posters (top 30) (2) the mean absolute difference between estimated and true scores of the truly best posters, and (3) the mean standard error of the population marginal mean estimator. Similar to the winning probability measure, we take the difference between NB1 and random, and NB1 and NB2, for each of the three measures in each iteration and summarize the distribution of the differences in Figure \ref{fig:other}. All three measures favor smaller values. We can see that the differences between NB1 and random are all significantly smaller than 0, which indicates NB1 is a better design than random assignment. The most obvious effect of having a near-balanced design is in reducing the standard error of the estimates. In turn, it will improve the estimation and ranking accuracy. Therefore, we see that the most significant difference between NB1 and random is in the standard error difference (3rd plot in the top row). Like winning probability, NB1 and NB2 are not significantly different in the three measures.

\section{Conclusion}
Motivated by a real example, we propose an algorithm to create a near-balanced incomplete block design that satisfies several conditions and to use a mixed effects model to estimate the poster scores. We propose two versions of the NB design, one balancing the number of reviews per poster (NB2) and the other one also balancing the reviews per pair of posters (NB1). We evaluate the accuracy and efficiency of the proposed design and demonstrate the benefits of balancing the number of reviews per poster over a random assignment. The simulation setup mimics the historical data in our example. Even though theoretically NB1 should outperform NB2 due to the additional balance, under this specific setting, the two versions of the NB design showed no significant performance differences for the four performance measures we considered, which implies that there is little difference in balancing the reviews per pair of posters. In some other settings, for example, when the poster, judge, and random errors are all smaller, it is possible that NB1 performs slightly better than NB2 (see Figures \ref{fig:win_555} and \ref{fig:other_555} in Appendix). In general, NB1 takes longer than NB2 to run and NB1 might not be feasible in some judge, poster, and review number combinations. Therefore, for simplicity and consistency, we would recommend NB2 to administrative practitioners.

We also discuss a general limitation of such judging systems by examining the proportion of the truly best posters receiving an award. We see that on average, only 60\% of the truly best posters will win an award. This is largely due to the limited reviews per judge and the assumption that the truly best 30 posters are not dramatically different from many of the other posters in our simulation study. Hopefully, this result will also offer some solace to those who don't win anything. Our recommendation has been accepted by the Graduate School for future poster competitions.  

The design proposed in this article can be generalized to many other settings such as consumer scoring surveys, website and customer service feedback surveys, and food tasting trials with a large number of samples. The code to generate the proposed design is available at: \url{https://psu.box.com/s/av1kekt1butboo3a2jrk79tmqqcr6myh}. 

\bibliographystyle{Chicago}
\bibliography{ref}

\newpage

\appendix

\section{Supplemental Figures}

\begin{figure}[!ht]
\hspace{-0.5cm}
\includegraphics[height=4in]{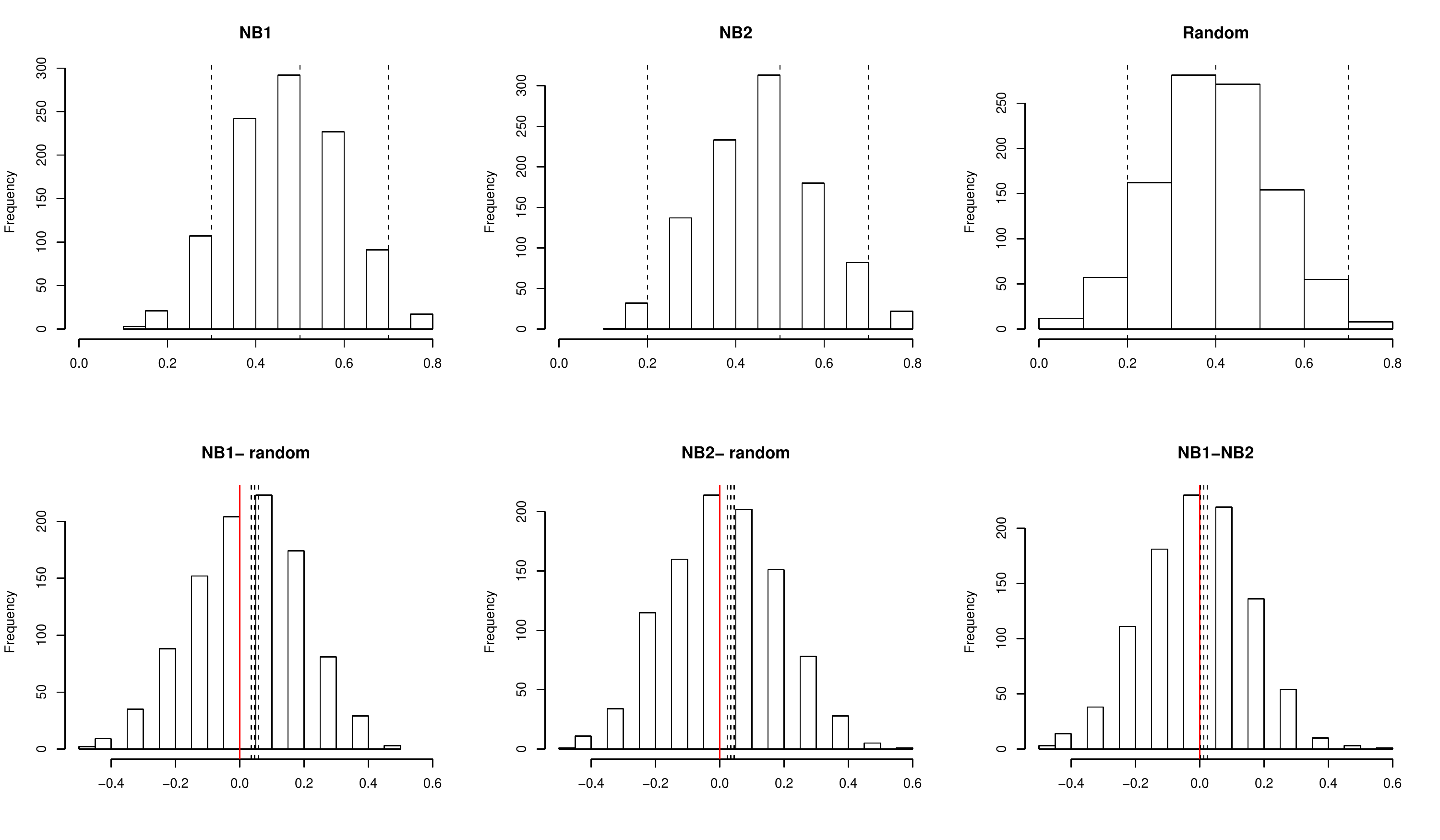}
\caption{Proportion of the 30 truly best posters judged to be in the top 30 when the standard deviations are (5,5,5) in Equation \ref{eqn:model}. The top row summarizes the distribution of the proportions of the three designs (NB1, NB2, and random) across the 1000 simulation runs. The vertical lines indicate the 2.5\%, 50\%, and 97.5\% quantiles of the distribution. The bottom row summarizes the distribution of the differences of the proportions between each pair of the designs. The solid vertical lines indicate 0. The two dashed vertical lines are the 95\% confidence bounds of the mean difference.}
\label{fig:win_555}
\end{figure}

\begin{figure}[!ht]
\hspace{-0.5cm}
\includegraphics[height=4in]{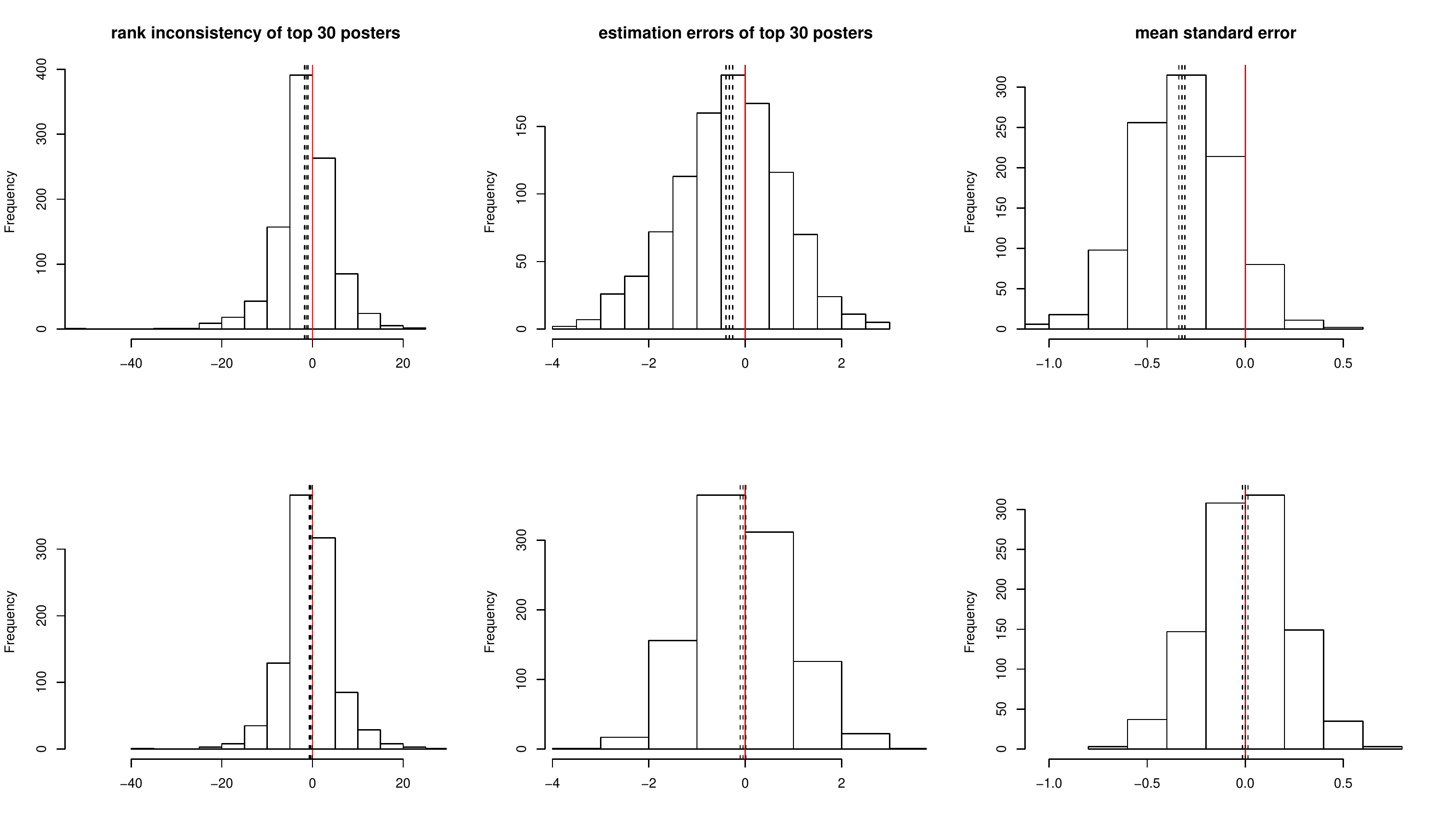}
\caption{Summary of the differences when the standard deviations are (5,5,5) in Equation \ref{eqn:model}, between NB1 and random assignment (top row) and NB1 and NB2 (bottom row) of three measures: median rank difference of the top 30 posters (first column), mean score deviance of the top 30 posters (2nd column), and mean standard error of the estimator (3rd column). The solid vertical line indicates 0. The dashed lines display the 95\% confidence bounds.}
\label{fig:other_555}
\end{figure}

\end{document}